\documentclass[preprint,review,11pt]{elsarticle}
\usepackage{amssymb,amsmath}
\usepackage{graphicx}
\usepackage{epstopdf}
\usepackage{epsfig}
\usepackage{color}
\usepackage{lineno}
\usepackage{subcaption}
\usepackage{siunitx}
\usepackage{natbib}
\usepackage{lineno}
\usepackage{lscape}
\usepackage{multirow}

\usepackage[version=3]{mhchem}

\usepackage[figuresright]{rotating}

\usepackage[colorlinks=true,
            linkcolor=red,
            urlcolor=blue,
            citecolor=gray]{hyperref}

\journal{Atmospheric Pollution Research}

\begin{document}


\begin{frontmatter}



\title{Multifractal characterisation of particulate matter ($PM_{10}$) time series in the Caribbean basin using visibility graphs}

\author[ks,ua]{Thomas Plocoste\corref{cor1}}
\ead{karusphere@gmail.com}

\author[cg]{Rafael Carmona-Cabezas}
\ead{f12carcr@uco.es}

\author[cg]{Francisco Jos{\'e} Jiménez-Hornero}
\ead{fjhornero@uco.es}

\author[cg]{Eduardo Guti{\'e}rrez de Rav{\'e}}
\ead{eduardo@uco.es}

\author[ua]{Rudy Calif}
\ead{rudy.calif@univ-antilles.fr}

\cortext[cor1]{Corresponding author}

\address[ks]{Department of Research in Geoscience, KaruSphère SASU, Abymes 97139, Guadeloupe (F.W.I.), France}

\address[ua]{Univ Antilles, LaRGE Laboratoire de Recherche en Géosciences et Energies (EA 4935), F-97100 Pointe-à-Pitre, France}

\address[cg]{Complex Geometry, Patterns and Scaling in Natural and Human Phenomena (GEPENA) Research Group, University of Cordoba, Gregor Mendel Building (3rd ﬂoor), Campus Rabanales, 14071, Cordoba, Spain}


\begin{abstract}
A good knowledge of pollutant time series behavior is fundamental to elaborate strategies and construct tools to protect human health. In Caribbean area, air quality is frequently deteriorated by the transport of African dust. In the literature, it is well known that exposure to particulate matter with an aerodynamic diameter of 10 $\mu$m or less ($PM_{10}$) have many adverse health effects as respiratory and cardiovascular diseases. To our knowledge, no study has yet performed an analysis of $PM_{10}$ time series using complex network framework. In this study, the so-called Visibility Graph (VG) method is used to describe $PM_{10}$ dynamics in Guadeloupe archipelago with a database of 11 years. Firstly, the fractal nature of $PM_{10}$ time series is highlighted using degree distribution for all data, low dust season (October to April) and high dust season (May to September). Thereafter, a profound description of $PM_{10}$ time series dynamics is made using multifractal analysis through two approaches, i.e. Rényi and singularity spectra. Achieved results are consistent with $PM_{10}$ behavior in the Caribbean basin. Both methods showed a higher multifractality degree during the low dust season. In addition, multifractal parameters exhibited that the low dust season has the higher recurrence and the lower uniformity degrees. Lastly, centrality measures (degree, closeness and betweenness) highlighted $PM_{10}$ dynamics through the year with a decay of centrality values during the high dust season. To conclude, all these results clearly showed that VG is a robust tool to describe times series properties.

\end{abstract}

\end{frontmatter}
%

\section{Introduction}	
\label{intro}

	Every year, millions of tons of mineral dust are transported from the African continent towards the Atlantic Ocean \citep{schutz1980, kaufman2005, koren2006, van2016}. According to literature, one of the most intense source of dust in the world is the Bodélé depression of the northern Lake Chad Basin \citep{choobari2014, euphrasie2020}. When dust sources are activated, main quantities of emitted dust can be lifted up in high altitudes and transported in the Saharan Air Layer (SAL) towards the Caribbean basin \citep{cavalieri2010, shao2011, schepanski2018}. The SAL is an hot and dry dust-laden layer confined between two inversion layers, i.e. at 1 and 5 $km$ height \citep{carlson1972, prospero1972, adams2012, tsamalis2013}. Because of intertropical convergence zone latitudinal movement through the year, a seasonal behavior is observed in dust cloud over the Atlantic Ocean \citep{nicholson2000}, moving  southward ($0-10^\circ$N) in winter and northward ($10-20^\circ$N) in summer \citep{moulin1997, holz2004, adams2012}.
	
	In Caribbean region, air quality is highly deteriorated by the seasonal transport of African dust \citep{prospero2014}. Contrary to urban areas highly industrialized in Europe, the United States or China, Caribbean islands exhibit low emission of particulate matter linked to anthropogenic pollution \citep{euphrasie2017, plocoste2017, plocoste2018}. After the long-range transport from African coast to the Caribbean, i.e. $\approx$ 5-7 days for 4000 $km$ \citep{velasco2018}, dust cloud is a mixture of fine and coarse particles \citep{petit2005, van2016}. The coarse particles (diameter $>$ 70 $\mu$m) are expected to deposit within less than one day \citep{mahowald2014, schepanski2018} while fine particles (diameter $<$ 10 $\mu$m, $PM_{10}$) may travelled throughout the world \citep{mahowald2014}. This study focused on $PM_{10}$ which are predominant in dust cloud \citep{petit2005}.
	
	In the literature, it is well known that $PM_{10}$ exposure has many adverse health effects as respiratory and cardiovascular diseases \citep{cadelis2014, gurung2017, zhang2017, momtazan2019}. By relating specific respiratory and cardiovascular emergency department admissions, \cite{feng2019} study shows that elderly (age $\geq$ 65 years) and male subjects are more susceptible to specific respiratory diseases. In addition, $PM_{10}$ exposure has also been associated with increased hospitalizations \citep{pun2014, su2016}. For these reasons, it is crucial to understand $PM_{10}$ dynamics in order to elaborate strategies and construct tools to protect Caribbean population. 
	
	Over the past few decades, numerous studies have shown complex networks efficiency to treat many topics \citep{amaral2004, costa2007, amancio2008, stam2010, gan2014, zou2019}, to cite a few. Usually, a complex network is used to characterize physical processes in nature and can be described as a graph showing non-trivial topological properties. During the last years, many methodologies have been developed in order to transform non-linear time series into complex networks \citep{zhang2006, lacasa2008, zou2019}, in order to take advantage of the many tools developed for analyzing the last ones. Thus, the so-called Visibility Graph (VG) algorithm introduced by \cite{lacasa2008} is performed here. In several environmental time series \citep{donner2012, pierini2012, carmona2020a}, this method has already shown that it perfectly inherits the main characteristics of the original time series \citep{lacasa2009, lacasa2010}.
	
	In VG frame, two approaches have been frequently used to analyze time series properties, i.e. multifractal analysis and centrality measures \citep{donner2012, carmona2019a, carmona2019b}. Contrary to traditional descriptive statistics which are only based on a single scale analysis, multifractal method assesses raw data over a wider range of temporal scales and exponents order \citep{zeleke2006}. In addition, multifractaly is defined by self-similarity properties, i.e. they are scale independent \citep{mandelbrot1982}. As regards centrality measures, several studies have shown that nodes distribution can be a useful way to describe time series dynamics \citep{donner2012, carmona2020b}. Here, the aim of this study is to perform a profound analysis of $PM_{10}$ concentrations in Guadeloupe archipelago according to African dust seasonality. To our knowledge, no study has yet investigated $PM_{10}$ dynamics in VG frame.
		
	The remainder of this paper is organized as follows. Section \ref{sitedata} depicts the study area and the experimental data. Section \ref{method} describes the theoretical framework. Section \ref{results} presents the results and discusses them. Finally, Section \ref{conclusion} gives the conclusion and an outlook for future studies.

\section{Study area and experimental data}	
\label{sitedata}

\begin{figure}[h!]
\begin{center}
\includegraphics[scale=1.0]{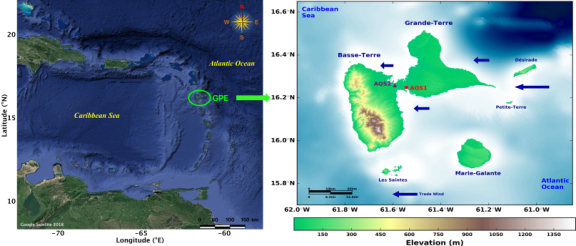}
\caption{\label{map} (a) shows an overview of the Caribbean basin with Guadeloupe archipelago located in the middle of the West Indies arc  ($16.25^\circ$N, $-61.58^\circ$W; GPE in green). (b) is a zoom of Guadeloupe archipelago with the locations of the Air Quality stations at Pointe-\`a-Pitre (AQS1, in a red circle) and Baie-Mahault (AQS2, in a purple triangle). The blue arrows indicate Trade winds direction.}
\end{center}
\end{figure}

Located in the middle of the Caribbean basin (see Figure \ref{map}(a)), Guadeloupe archipelago ($16.25^\circ$N $-61.58^\circ$W, GPE) is a French overseas region of 390,250 inhabitants which cover an area of $\sim$1800 $km^2$ \citep{plocoste2019a}. Due to its geographical position, GPE experiences a tropical climate with uniformed temperature and continuous high humidity \citep{plocoste2014}. Hourly $PM_{10}$ measurements are released by GPE air quality network which is managed by Gwad'Air agency (http://www.gwadair.fr/). The Air Quality Stations (AQS) are principally at the center of the island where is located Guadeloupe's economic lung. A quarter of the population lives there and the topography is nearly flat (green space in Figure \ref{map}(b)). In order to study the impact of large scale events on $PM_{10}$ fluctuations, hourly data is converted into daily average values. From 2005 to 2017, only one $PM_{10}$ sensor was available on this air quality network. It was successively placed in Pointe-à-Pitre (AQS1, $16.2422^\circ$N $61.5414^\circ$W) from 2005 to 2012 and Baie-Mahault (AQS2, $16.2561^\circ$N $61.5903^\circ$W) from 2015 to nowadays. Due to their geographical proximity ($\sim$8.1 $km$), both measurements are made under the same environmental conditions. Consequently, 11 years of $PM_{10}$ data are available to conduct this study, i.e. 3849 points. 

\begin{figure}[h!]
\centering
\includegraphics[scale=0.45]{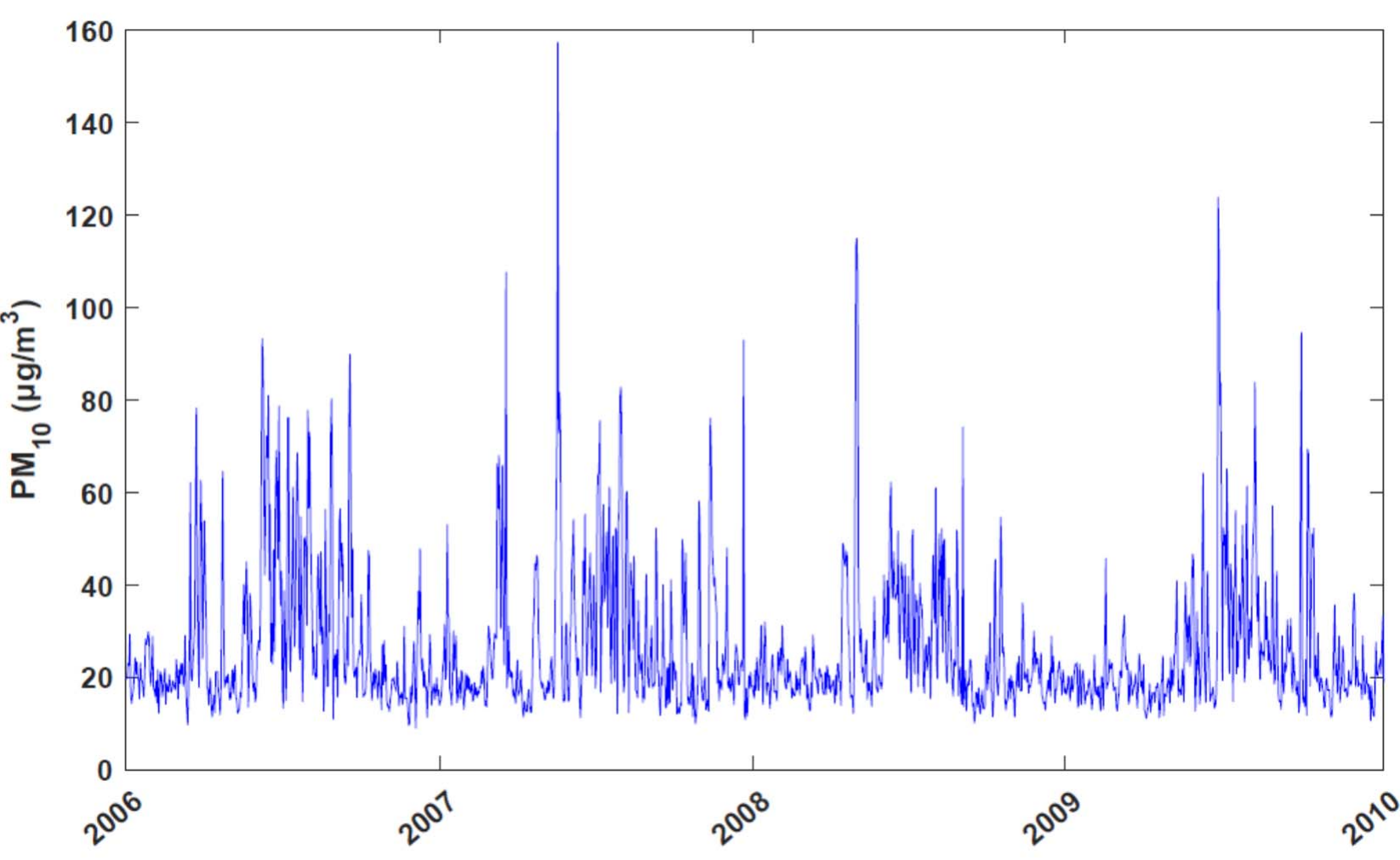}
\caption{\label{signal} A sequence of daily average data of $PM_{10}$ concentrations between 2006 and 2009.}
\end{figure}

Figure \ref{signal} shows a sequence of $PM_{10}$ daily signal analyzed in this work illustrating huge fluctuations. The strong oscillations observed in the middle of each year are attributed to $PM_{10}$ related to dust outbreaks coming from the African coast from May to September, i.e. the high dust season \citep{plocoste2020}. From October to April (low dust season), the $PM_{10}$ concentrations are mainly linked to anthropogenic activity and marine aerosols \citep{euphrasie2020}. In order to analyze the seasonal behavior of $ PM_ {10} $ data, the signals are reconstructed for the low season (2253 points) and the high season (1596 points) over 11 years.\\

\section{Theoretical framework}	
\label{method}

\subsection{Visibility graphs}
\label{vgmetho}

	In graph theory, a graph is defined as a set of vertices, points or nodes connected to each other by lines that are usually called edges. Visibility Graph (VG) is a robust tool that transform time series into a graph. Firstly introduced by \cite{lacasa2008}, this new complex network has the benefit to inherit many properties of the original time series. 

	In order to obtain the VG of a time series, it is essential to establish a criterion to determine which points (or nodes) are linked to each other, i.e. have visibility properties. Consequently, let $y(t)$ be a function of time (time series), two arbitrary points ($t_a$, $y_a$) and ($t_b$, $y_b$) will have visibility, i.e. will be connected in the graph, if any given point ($t_c$, $y_c$) between them ($t_a < t_c < t_b$) meets the following condition \citep{lacasa2008}:

\begin{equation}
y_{c} < y_{a} + (y_{b}-y_{a}) \frac{t_{c}-t_{a}}{t_{b}-t_{a}}
\label{VG}
\end{equation}

\begin{figure}[h!]
\centering
\includegraphics[scale=0.7]{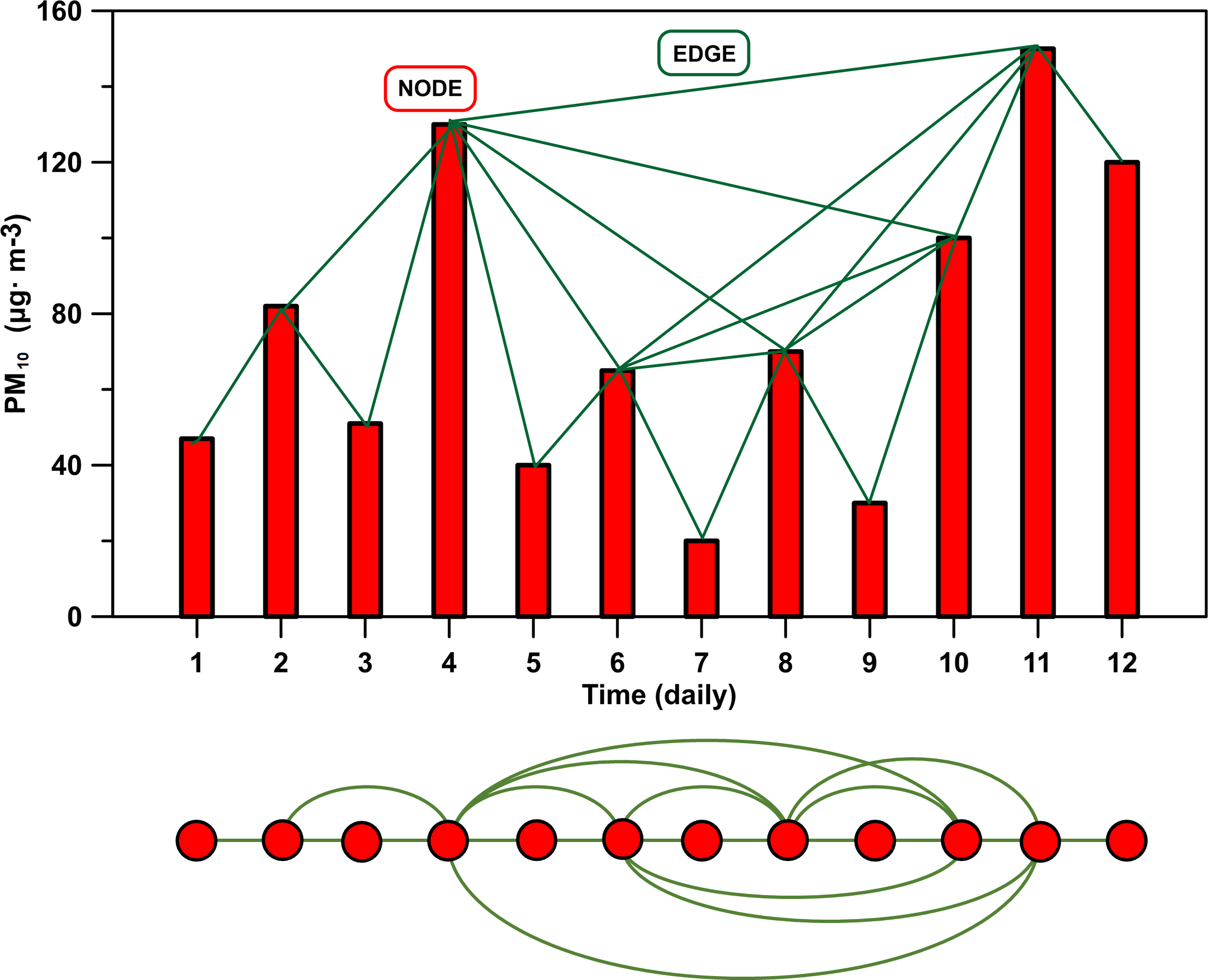}
\caption{\label{visibility} Sample of $PM_{10}$ time series transformed into a complex network through the visibility algorithm. The nodes of the graph are the data points (red bars), while the links among them (edges) are illustrated as solid lines. At the bottom, all the connections are illustrated in a more visual way.}
\end{figure}

	After this condition for every pair of points in the series is checked, it is possible to determine which pairs have visibility or not. As an example, Figure \ref{visibility} illustrates the transformation of a time series into a complex network through VG frame. 
	
	By performing visibility method, a N$\times$N adjacency binary matrix is achieved with N the total number of points in the studied time series. The information of the nodes is given by each row of the matrix. Thus, when $a_{ij} = 1$ the node $i$ and $j$ have visibility while $a_{ij} = 0$ means that no edge connects those two nodes, i.e. no visibility. In order to simplify the algorithm and reduce the computational required time, the resulting matrix exhibits several properties \citep{carmona2019a}. Firstly, all the elements in the diagonal are zero ($a_{ij} = 0$). Indeed there is no visibility of an element with itself since there are no intermediate points to meets the criterion. Secondly, the condition of visibility is reciprocal, i.e. if a node “sees” another, this other “sees” the first one as well. This is translated into the adjacency matrix being symmetric: $a_{ij} = a_{ji}$. Thirdly, the elements surrounding the diagonal are always 1 ($a_{ij} =1$ for $j = i \pm 1  $) since each point always sees the closest previous and next node. When taking all of that into consideration, the visibility $V$ matrix can be described as below \citep{carmona2019a}:   
	
\begin{equation}
V= 
 \begin{pmatrix}
  0 & 1 & \cdots & a_{1,N} \\
  1 & 0 & 1 & \vdots \\
  \vdots  & 1  & \ddots & 1 \\
  a_{N,1} & \cdots & 1 & 0
 \end{pmatrix}
\label{Mat}
\end{equation}

The degree of a node ($k_i$) is the number nodes that have reciprocal visibility with the first one. As an example in Figure \ref{visibility}, the degree of the first node is $k=1$, for the second one $k=3$, for the third one $k=2$, etc. By taking into account the degree of each one of the nodes present in the VG,  the degree distribution of the sample ($P(k)$) can be computed. In literature, many studies have highlighted the efficiency of VG method to describe the nature of a time series \citep{lacasa2008, mali2018, carmona2019b}. Consequently, we can firstly overview the behavior of $PM_{10}$ time series by studying the degree distribution before performing more complex multifractal analysis. According to \cite{lacasa2009}, \cite{lacasa2010} and \cite{mali2018}, scale free VG (which arise from fractal time series) have degree distributions that can be fitted to power laws such as $P(k) \propto  k^{-\gamma}$. This is due to the so-called effect of hub repulsion \citep{song2006}.

\subsection{Multifractal analysis}
\label{methodMulti}

	In order to investigate more precisely the dynamics of a time series, it is essential to analyze more scaling exponents, including through fractal and multifractal analysis \citep{pamula2014}. Multifractal approach is regarded as the inherent property of complex and composite systems because it gives the possibility of having different densities depending on the region of application \citep{mandelbrot1974}. In the literature, there are two ways of depicting multifractality behavior: the generalized fractal dimension $D_q$ \citep{tel1989, block1990, schreiber1991, posadas2001} and the singularity spectrum $f(\alpha)$ \citep{chhabra1989, bacry1993, lyra1998, caniego2005}. Multifractal analysis has been widely carried out using the fixed-size algorithms (FSA) \citep{halsey1986, mach1995}. In this study, the sandbox algorithm is applied in complex network frame as \cite{liu2015} and \cite{mali2018}.

\subsubsection{Generalized fractal dimension}
\label{methodGen}

	The first method introduced to study multifractal formalism are the generalized fractal or Rényi dimensions $D_q$ \citep{harte2001}. This latter depicts the scaling exponents of the $qth$ moments of the system and can be defined by \citep{feder1988}: 

\begin{equation}
D_q = \frac{1}{q-1}\, \lim_{\delta\to0} \frac{lnZ_q(\delta)}{ln\,\delta}\,\forall q \neq1
\label{gen1}
\end{equation}

\begin{equation}
D_1 = \lim_{\delta\to0} \frac{\sum^{N_c(\delta)}_{i=1}\,\mu_i(\delta)\,ln\mu_i(\delta)}{ln\,\delta}
\label{gen2}
\end{equation}

where $q$ is the moment order, $\delta$ is the size of the used cells to cover the
system, $Z_q(\delta)$ is the partition function, $N_c(\delta)$ is the number of cells with length $\delta$ and $\mu_i(\delta)$ is the probability measurement of each cell. By taking the limit of $D_q$ when $q \to 1$, $D_1$ expression is achieved. The partition function previously introduced, is defined as follows \citep{liu2015}:

\begin{equation}
Z_q(\delta) = \sum^{N_c(\delta)}_{i=1}\,[\mu_i(\delta)]^q
\label{gen3}
\end{equation}

From Equation \ref{gen1}, several parameters can be obtained \citep{carmona2019b}: i) $D_{q=0}$ is the fractal dimension of the given system or box-counting dimension; ii) $D_{q=1}$ corresponds to the so-called information entropy; iii) $D_{q=2}$  is the correlation dimension. $Dq$ limits for $q$ values ranging between -$\infty$ and +$\infty$, depict the scaling properties of the regions where the regions are respectively more rarefied and concentrated. Usually, the strength of the multifractality can be quantified by $\Delta\,D_q = max\,D_q - min\,D_q$ \citep{yu2016}.   

\subsubsection{Singularity spectrum}
\label{methodSing}

	Another way to assess the multifractal properties of a time series is the singularity or multifractal spectrum. Traditionally, the Legendre transformation from mass exponents $\tau(q)$ is used to determine it \citep{muzy1993}. Nevertheless, several authors as \cite{chhabra1989} and \cite{veneziano1995} pointed out several drawbacks and errors in this method due to the inclusion of spurious points and error amplification from the derivative. In addition, the Legendre transform does not allow independent measurements of the Rényi spectrum as $\tau(q) = (1 - q)D_q$. To overcome this drawbacks, \cite{chhabra1989} introduced a method to determine the $\alpha$-spectrum directly from the original time series. In order to compute the probabilities of the boxes of radius $r$, this method is based on the normalized measure $\beta_i(q)$ and $\mu_i$ from the original time series with \citep{chhabra1989}:     

\begin{equation}
\beta_i(q,r) = [P_i(r)]^q\,/\,\sum_{j}[P_j(r)]^q
\label{sing1}
\end{equation}

where $P_i(r)$ are the different fractal measurements for each box of radius $r$, i.e. the number of nodes. Thus, from Equation \ref{sing1}, $f(\alpha)$ and $\alpha$ can be computed by using the following equations \citep{chhabra1989}: 

\begin{equation}
f(q) = \lim_{r\to0} \frac{\sum_{i}\,\beta_i(q,r)\,log[\beta_i(q,r)]}{log\,r}
\label{sing2}
\end{equation}

\begin{equation}
\alpha(q) = \lim_{r\to0} \frac{\sum_{i}\,\beta_i(q,r)\,log[P_i(r)]}{log\,r}
\label{sing3}
\end{equation}

In the literature, $\alpha$ is termed the Lipschitz-Hölder exponent \citep{posadas2001}. Technically, those parameters are estimated using the slope of $\sum_{i}\,\beta_i(q,r)\,log[\beta_i(q,r)]$ over $log\,r$ for $f(\alpha)$; and $\sum_{i}\,\beta_i(q,r)\,log[P_i(r)]$ over $log\,r$ for $\alpha(q)$. This slope is characterized by means of a linear regression in the same range of radii where the other fractal measures are calculated \citep{carmona2019b}. As the generalized fractal dimension, the strength of the multifractality can be estimated by the width of the spectrum with $W = \alpha_{max} - \alpha_{min}$ \citep{mali2018}.

\subsubsection{Sandbox algorithm}
\label{methodSing}

	For the purpose of easily computing fractal dimensions of real data, \cite{tel1989} introduced a sandbox algorithm (SBA) which is originated from the box-counting algorithm \citep{halsey1986}. SBA was developed by \cite{vicsek1990} and firstly applied to multifractal analysis of complex networks by \cite{liu2015}. Contrary to other box-counting FSA, SBA is able to correctly estimate the side corresponding to the negative values of $q$ from both Rényi and singularity spectra. According to \cite{yu2016}, SBA is the most effective, feasible and accurate algorithm to investigate the multifractal behavior and compute the mass exponent of complex networks. 

	In more concrete terms, a number of randomly placed boxes are selected for each radius. These latter are always centred in a non-zero point of the system, i.e. a node. Consequently, the entire network is covered with those boxes by choosing a sufficiently high number of them \citep{carmona2019b}. The probability measurement used to calculate each box ($B$) is defined by the following equation \citep{liu2015}:
	
\begin{equation}
\mu(B) = \frac{M(B)}{M_0}
\label{SBA1}
\end{equation}

After computed that quantity for each sandbox of a given radius, the generalized fractal dimensions defined in Equation \ref{gen1} of section \ref{methodGen} can be improved by the following formula \citep{liu2015}: 

\begin{equation}
D_{q}^{sb} = \frac{1}{q-1}\, \lim_{r\to0} \frac{ln\left<\mu(B)^{(q-1)}\right>}{ln\,r}\,\forall q \neq1
\label{SBA2}
\end{equation}

For $q = 1$, SBA can also adjust to Equation \ref{gen2} with \citep{mali2018}:

\begin{equation}
D_{1}^{sb} = \lim_{r\to0} \frac{<ln\mu(B)>}{ln\,r}
\label{SBA3}
\end{equation}

The SBA procedure for multifractal analysis of complex networks is well described by \cite{liu2015}, \cite{yu2016} and \cite{carmona2019b}. In this study, the following parameters are used to perform the SBA: i) the interval used for the radii goes from 1 to 30 ($r \in [1, 30]$) according to the distance matrix between the nodes; ii) the range of moments is between $q = -5$ and $q = +5$ with an increment step of 0.25.

\subsection{Centrality measures}
\label{centralitymetho}

	In VG frame, it is essential to characterize nodes importance of the complex networks.  There are different measures to quantified the relevance of individual vertices for structural as well as dynamic characteristics of the underlying system \citep{donner2012}. To achieve this, the centrality measures are frequently used. Firstly introduced by \cite{freeman1978} for social network study, it was later used in complex networks for many fields \citep{donner2012, natapov2013, carmona2020b}. Here, we focused on three centrality measures: degree, closeness and betweenness.

\subsubsection{Degree centrality}
\label{methodDC}

	In complex network, the degree centrality captures how many destinations can be seen from each node within given geometrical conditions. In other words, the degree of a node ($k_i$) can be defined as the number nodes that have reciprocal visibility with the first one. It is defined by the following equation \citep{donner2012}:

\begin{equation}
k_i = \sum_{j} a_{ij}
\label{deg}
\end{equation}

As a reminder, by taking Figure \ref{visibility} as an example, the degree of the three first node are respectively, $k_1=1$, $k_2=3$ and $k_3=2$. The higher the value of $k_i$, the greater the centrality is.

\subsubsection{Closeness centrality}
\label{methodClo}

	In order to define closeness centrality, it is important to introduce another main property of graph theory frame, the Shortest Path (SP). Contrary to degree centrality which takes into account the number of edges and the adjacency matrix properties,  closeness centrality estimates how many steps are required to access every other node from a given node. More concretely, there is a different number of edges and paths between two distant nodes $(i, j)$. However, there will be some of these paths where the number of edges will be minimum, i.e SP. For a node $i$, this quantity is defined by the inverse of the sum of distances from this node to others \citep{carmona2020b}:

\begin{equation}
c_i = \frac{1}{\sum^{N}_{j=1}\,d_{i,j}}
\label{clo}
\end{equation}

with $d_{i,j}$ the $(i, j)$ element of the distance matrix. Closeness centrality evaluates the relative importance of a node within the graph. The higher the value of $c_i$, the greater the nodes are integrated with SP from the others in the graph.

\subsubsection{Betweenness centrality}
\label{methodBet}

	As closeness centrality, betweenness centrality is also based on SP. It indicates how many times a node is passed through SP of other pairs of nodes. For a node $i$, betweenness centrality  is defined as follows \citep{carmona2020b}:   
	
\begin{equation}
  b_i =
  \sum^{N}_{\substack{j=1\\                 
                  j\neq i}}\\
  \sum^{N}_{\substack{k=1\\                 
                  k\neq i,j}}
        \frac{n_{jk}(i)}{2N_{jk}}
\label{bet}         
\end{equation}

with $n_{jk}(i)$ the number of SP's from $j$ to $k$ that include $i$ and $N_{jk}$ the total number of SP's between $j$ and $k$. Betweenness centrality estimated the influence a node has over the flow through the network. The higher the value of $b_i$, the greater the nodes weights are important in the graph. \\

\begin{figure}[h!]
\centering
\includegraphics[scale=0.45]{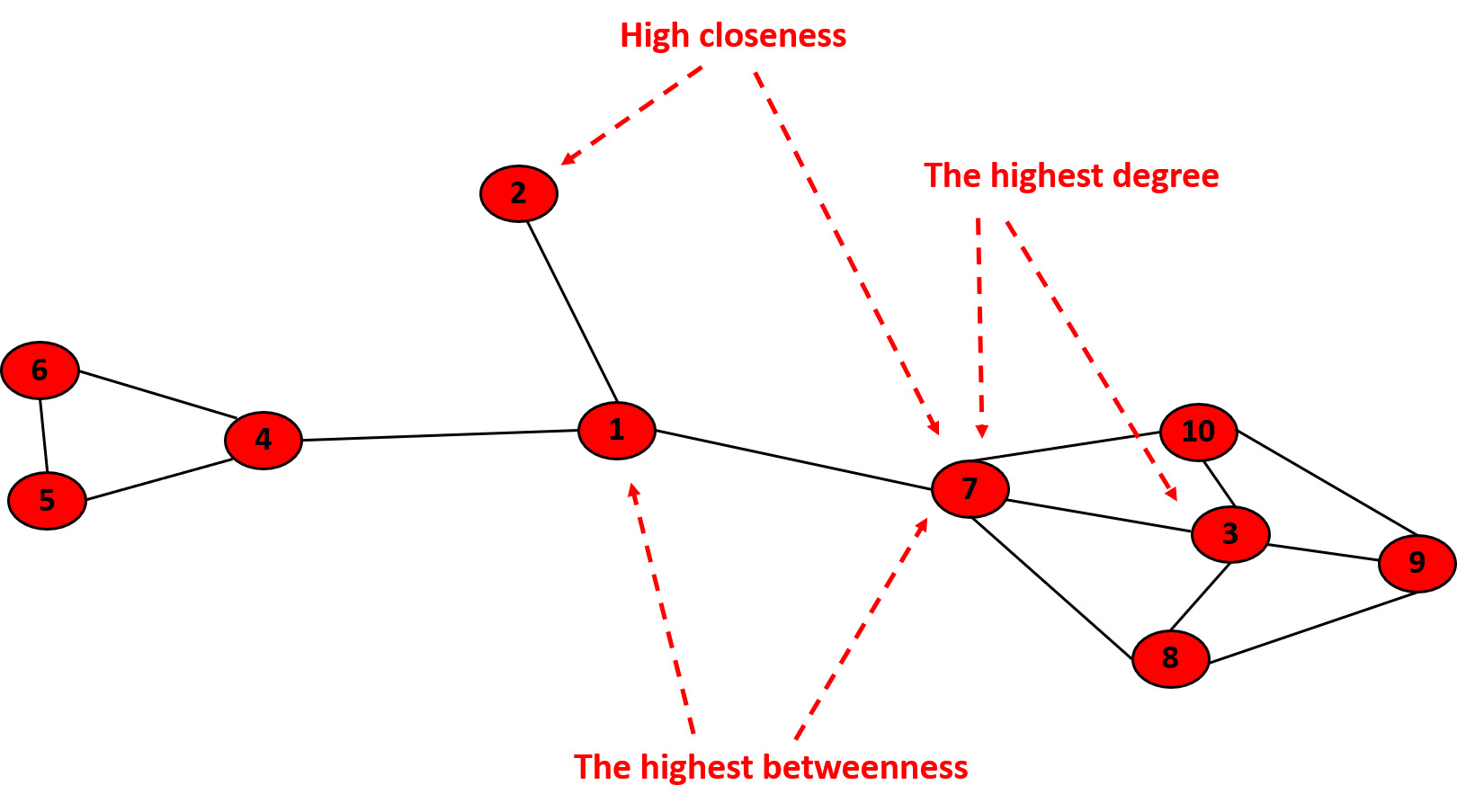}
\caption{\label{CentGraph} An example of centrality measures for 10 nodes in Integrative Visibility Graph \citep{natapov2013}.}
\end{figure}

	It is important to underline that a node with a high degree centrality does not guarantee that it is well connected to all other nodes. In some cases, a node with few direct edges is more important since it can act as a bridge, i.e. without it a network may be broken into subgraphs \citep{natapov2013}. Figure \ref{CentGraph} illustrates an example of these cases for 10 nodes based on \cite{natapov2013} study for Integrative Visibility Graph. Here, node 1 shows one of the highest betweenness while node 3 has the highest degree ($k = 4$). Node 7 combine both previous centrality measures an add high closeness, meaning that is the most integrated node in the graph. Regarding node 2, it exhibits the smallest possible betweenness and degree ($k = 1$), however it shows high closeness. Consequently, it may be considered as well integrated.

\section{Results and Discussion}
\label{results}

\subsection{Degree distribution}
\label{resultDeg}

	Before performing an in-depth analysis of $PM_{10}$ data, it is essential to firstly investigate the nature of this time series, i.e. periodic, random or fractal as an example. To achieve this, the degree distribution $P(k)$ of the VGs is analyzed. This quantity is defined as the number of nodes of a given degree divided by the total number of them in the VG. According to \cite{carmona2019b}, large degrees are related to nodes corresponding to highest concentrations of pollutants such as tropospheric ozone (so-called “hub”), they are much less likely to be repeat within the network. Consequently, hubs which traditionally have high visibility should correspond to $PM_{10}$ extreme values. In the Caribbean, $PM_{10}$ extreme values are mainly attributed to African dust outbreak \citep{plocoste2020}. In other words, hubs likeliness will be less significant than nodes close to $PM_{10}$ average. Further analysis for the relationship between hubs and empirical data will be carried out later on. 
	
\begin{figure}[h!]
\centering
\includegraphics[scale=0.25]{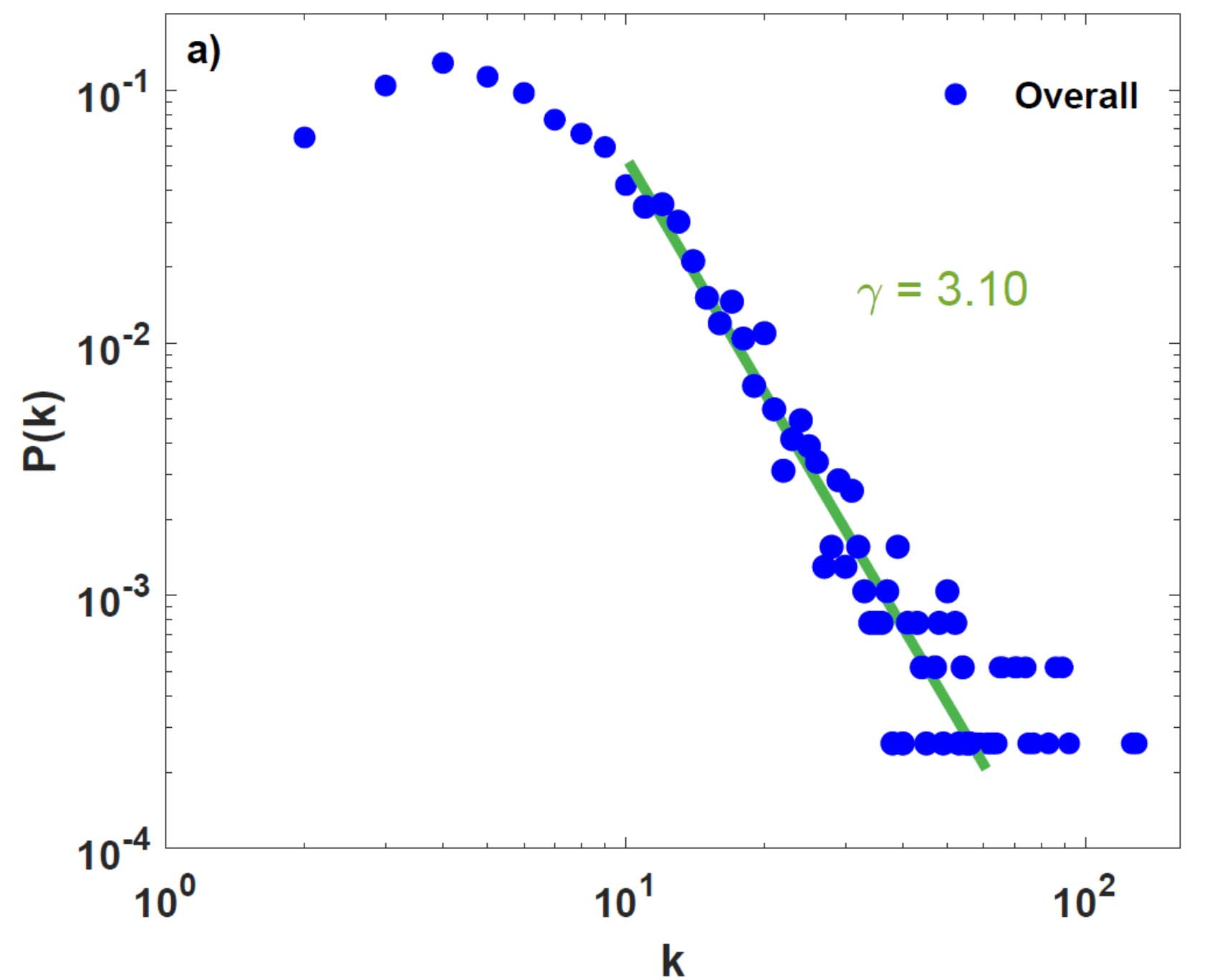}
\includegraphics[scale=0.25]{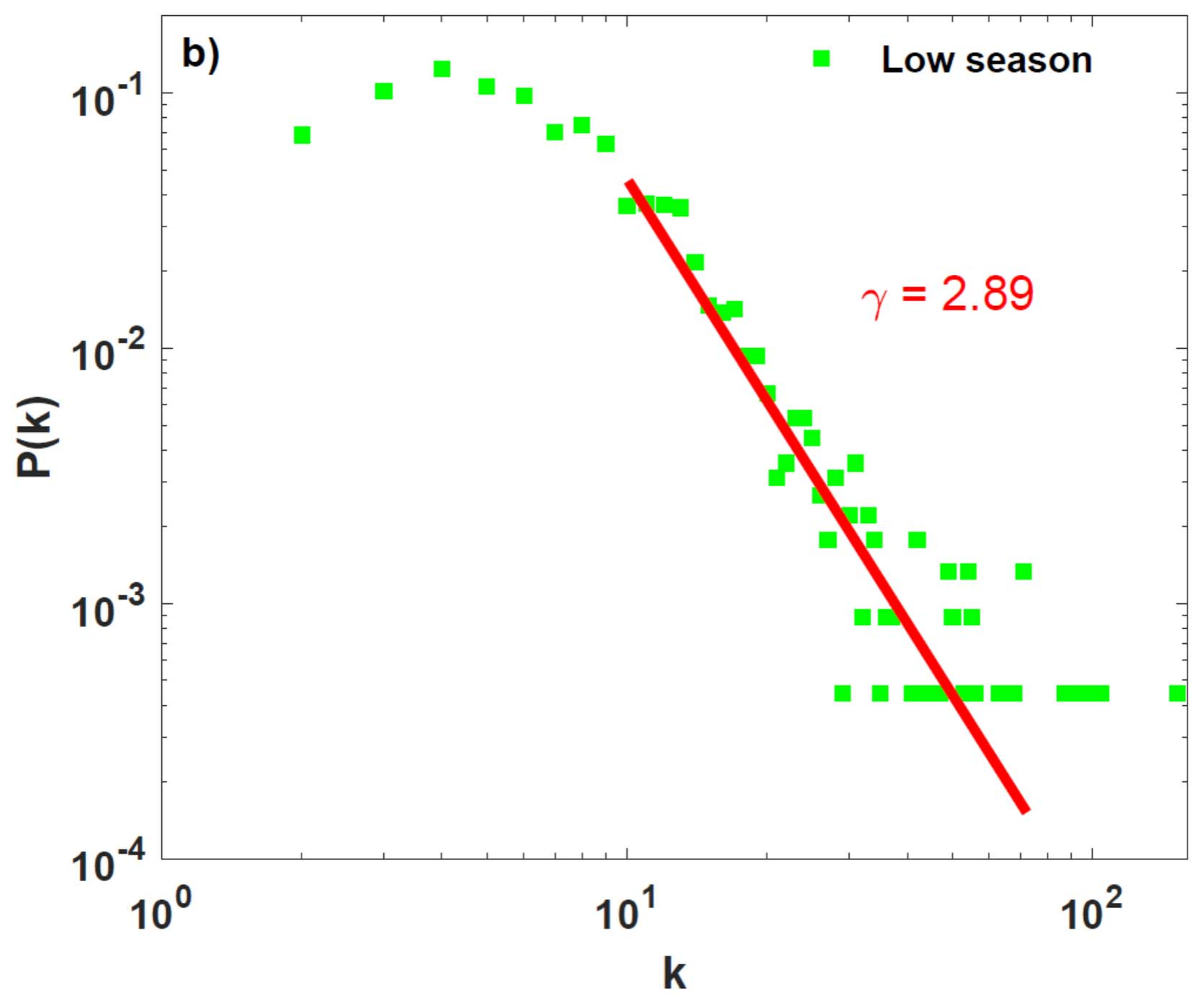}
\includegraphics[scale=0.25]{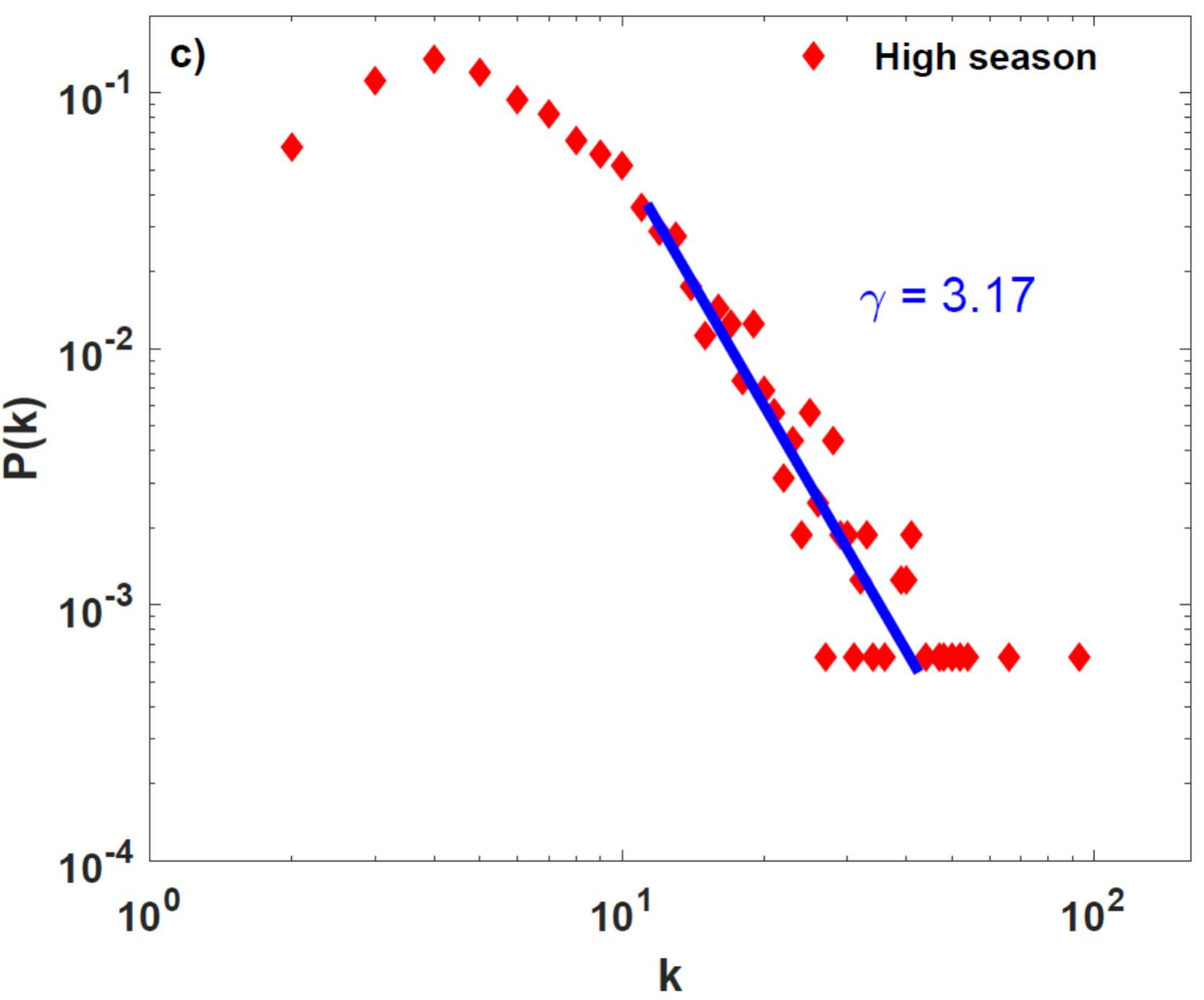}
\caption{\label{resDeg} Degree distribution of the visibility graph for (a) all data, (b) low dust season (October to April) and (c) high dust season (May to September) in log-log plot. As noted, the tail of the distribution of $PM_{10}$ time series exhibits a scale-free behavior because it can be fitted by a power law in all cases.}
\end{figure}

	Figure \ref{resDeg} illustrates the degree distribution obtained for all data, low dust season (October to April), and high dust season (May to September) over 11 years. For all cases, the tail region of the log-log plot of $P(k)$ can be fitted by a power law like $P(k) \propto  k^{-\gamma}$. In VG analysis,  the exponent $\gamma$ is relevant because it is a scale free parameter for a large number of real networks \citep{mali2018, carmona2019b}. Furthermore, this exponent can be related to the Hurst exponent in some cases \citep{lacasa2009, lacasa2010}. The exponents computed from the slopes are estimated for $k \geq 10$ with $2.89 \leqslant \gamma \leqslant 3.17$. It can be noticed that $\gamma$  value for the overall is centred between $\gamma$  values for low and high dust seasons. All this results clearly show the fractal behavior of $PM_{10}$ time series.	
	
	Classically, this approach is used to describe time series nature. However, for a profound description of $PM_{10}$ time series dynamics, a multifractal analysis is required.

\subsection{Multifractal analysis}
\label{resultdMulti}

\subsubsection{R{\'e}nyi spectrum}
\label{resultRen}

\begin{figure}[h!]
\centering
\includegraphics[scale=0.45]{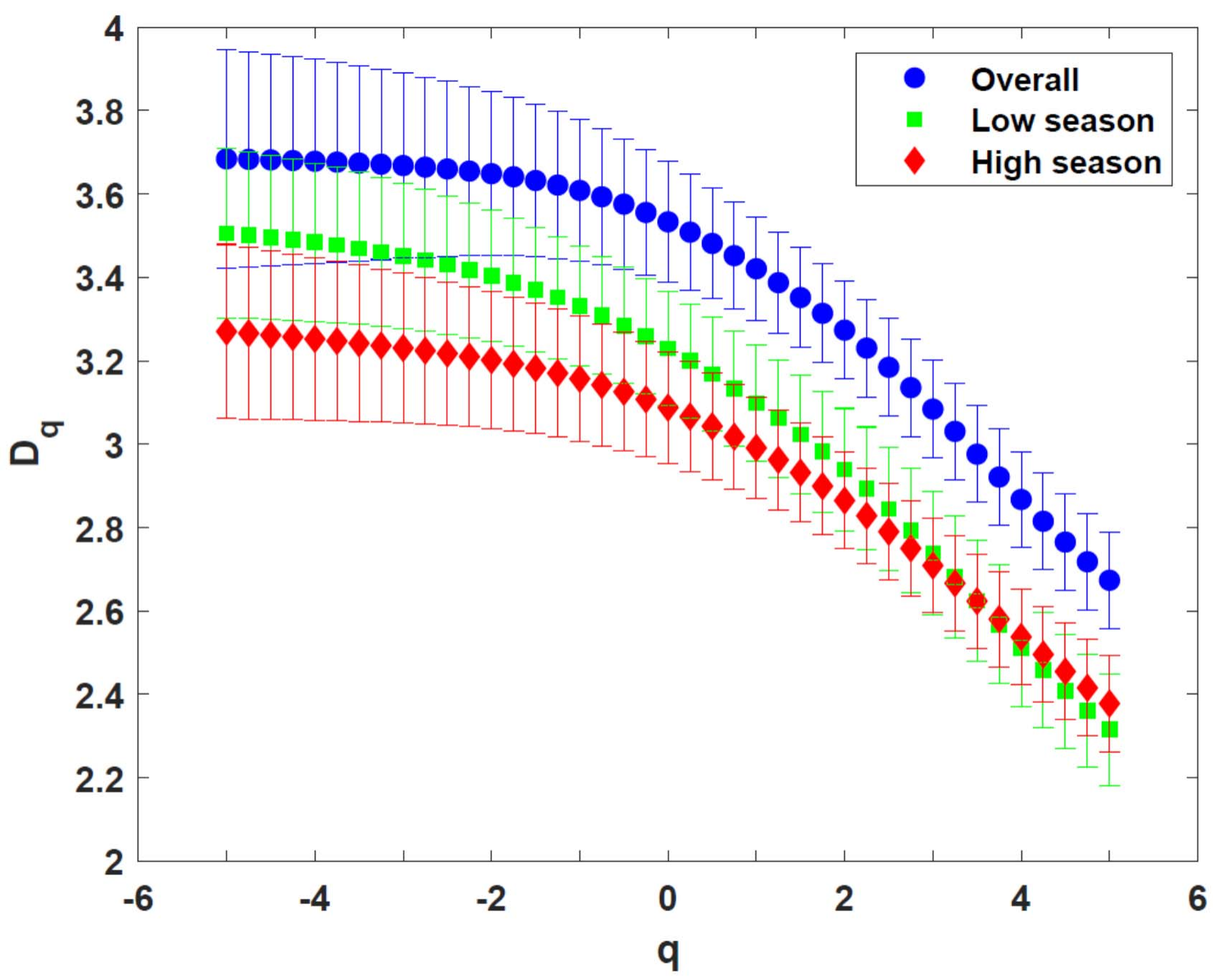}
\caption{\label{resReny} Rényi dimensions for all data, low dust season (October to April) and high dust season (May to September). Standard deviations are illustrated by the whiskers.}
\end{figure}

	In order to study the multifractal properties of $PM_{10}$ time series in VG frame with Rényi spectrum, the SBA procedure is firstly performed. Thus, the quantities $\frac{ln\left<\mu(B)^{(q-1)}\right>}{q-1}$ for $q \neq1$ and $<ln\mu(B)>$ for $q = 1$ are plotted against $ln\,r$. As the results of partition functions are widely presented in literature \citep{liu2015, yu2016, mali2018, carmona2019b}, we just recall here the range where the linear regression is applied for all $q$ values in order to build Rényi spectrum from Equation \ref{SBA2} and \ref{SBA3}. From $q = -5$ to $q = +5$ with an increment step of 0.25, a linear regression was made for $0 \leqslant ln\,r \leqslant 2$. Figure \ref{resReny} illustrates the Rényi spectra obtained where $D_q$ values are plotted against order number $q$. One can clearly observe the multifractal properties of $PM_{10}$ time series in the Caribbean area with VG procedure because $D_0 > D_1 > D_2$ for all cases (see Table \ref{VGresult}). This result validates those previously obtained by \cite{plocoste2017} with the classical structure function analysis. Traditionally, the strength of the multifractality $\Delta\,D_q$ can be quantified by the difference between the maximum and the minimum values of $D_q$ \citep{yu2016}. Table \ref{VGresult} shows that $\Delta\,D_q$ value is higher for low dust season, followed by “overall” and high dust season. These results are consistent with those previously found by \cite{plocoste2020a} with descriptive statistics. Indeed, by computing the kurtosis parameter ($K$) which is a useful indicator of intermittency in pollution studies \citep{windsor2001, plocoste2018}, they found $K_{Low}(33.0) > K_{Overall}(11.8) > K_{High}(5.9)$.

\begin{landscape}
\begin{table}[!p]
\small
\begin{tabular}{c |c c c| c c c c c c| c c c}
\hline
\multicolumn{1}{c}{}& \multicolumn{3}{|c|}{Degree distribution} & \multicolumn{6}{c|}{Rényi parameters}&\multicolumn{3}{c}{Singularity parameters}\\
\hline
\multicolumn{1}{c}{Period}& \multicolumn{1}{|c}{$\gamma$} & \multicolumn{1}{c}{$\bar{k}$}&\multicolumn{1}{c|}{$\sigma_k$}& \multicolumn{1}{c}{$D_0$} & \multicolumn{1}{c}{$D_1$}&\multicolumn{1}{c}{$D_2$}& \multicolumn{1}{c}{$D_0 - D_1$} & \multicolumn{1}{c}{$D_0 - D_2$}&\multicolumn{1}{c|}{$\Delta\,D_q$}& \multicolumn{1}{c}{$\alpha_{min}$} & \multicolumn{1}{c}{$\alpha_{max}$}&\multicolumn{1}{c}{$W$} \\
\hline
Overall & 3.10 & 65.5 & 37.10 & 3.5333 & 3.4209 & 3.2735 & 0.1124 & 0.2598 & 1.0106 & 2.663 & 3.761 & 1.098\\
\hline 
Low season & 2.89 & 77.0 & 43.73 & 3.2300 & 3.0986 & 2.9394 & 0.1314 & 0.2906 & 1.1899 & 2.293 & 3.616 & 1.323\\
\hline
High season & 3.17 & 47.0 & 26.99 & 3.0878 & 2.9911 & 2.8648 & 0.0967 & 0.2230 & 0.8924 & 2.282 & 3.350 & 1.068\\
\hline
\end{tabular}
\caption{\label{VGresult} Characteristics values from VG frame for all data, low dust season (October to April) and high dust season (May to September).}
\end{table}
\end{landscape}

\begin{figure}[h!]
\centering
\includegraphics[scale=0.25]{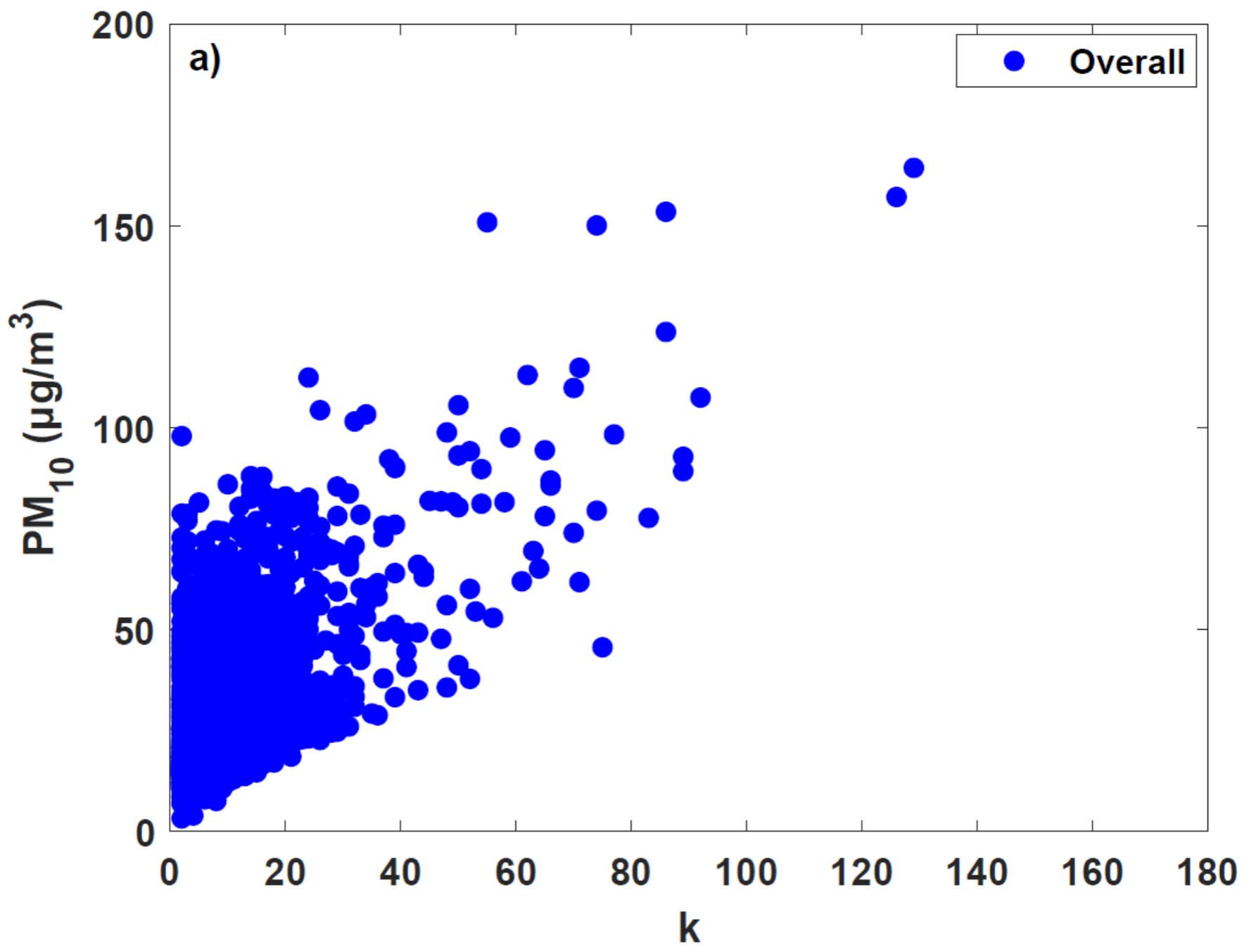}
\includegraphics[scale=0.25]{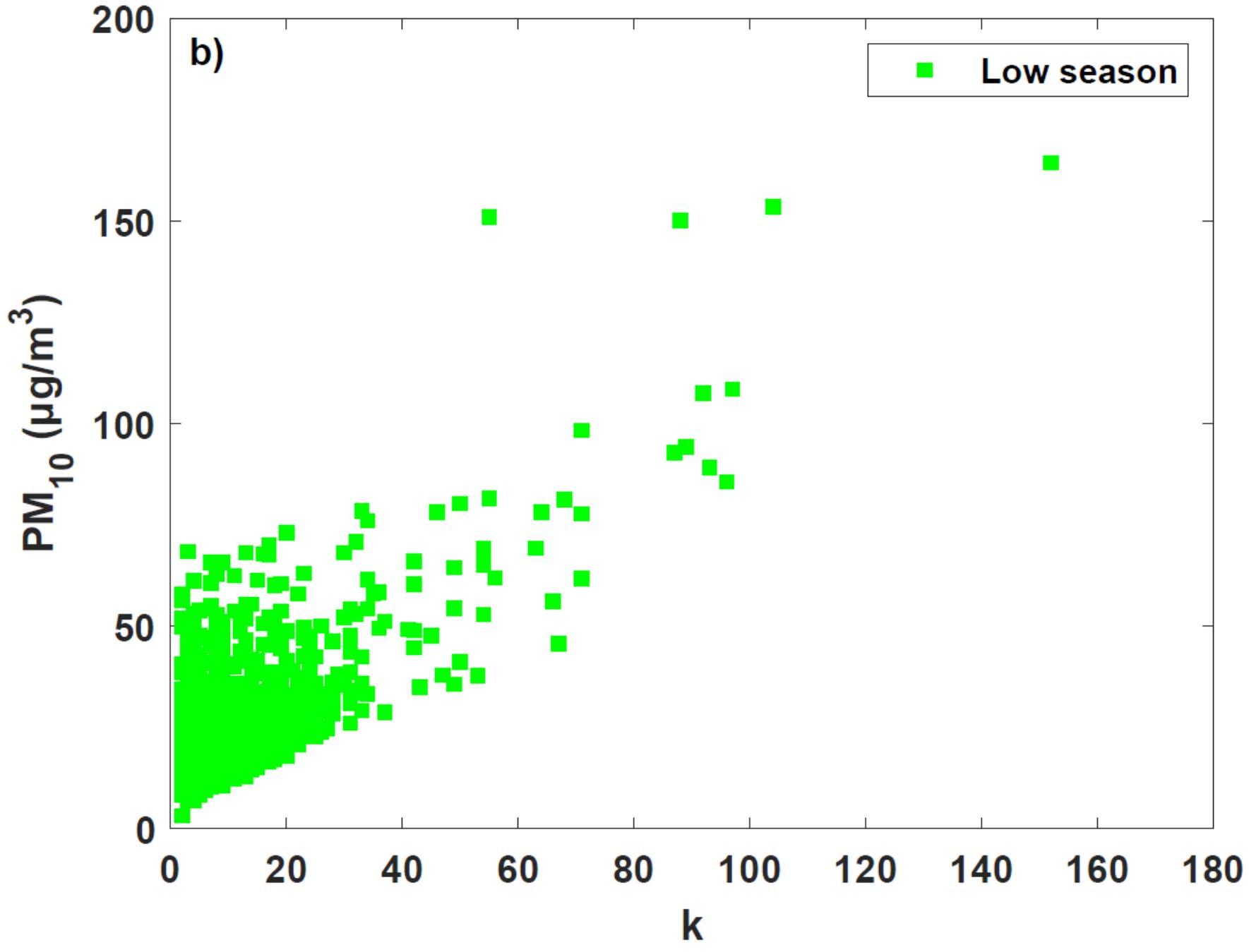}
\includegraphics[scale=0.25]{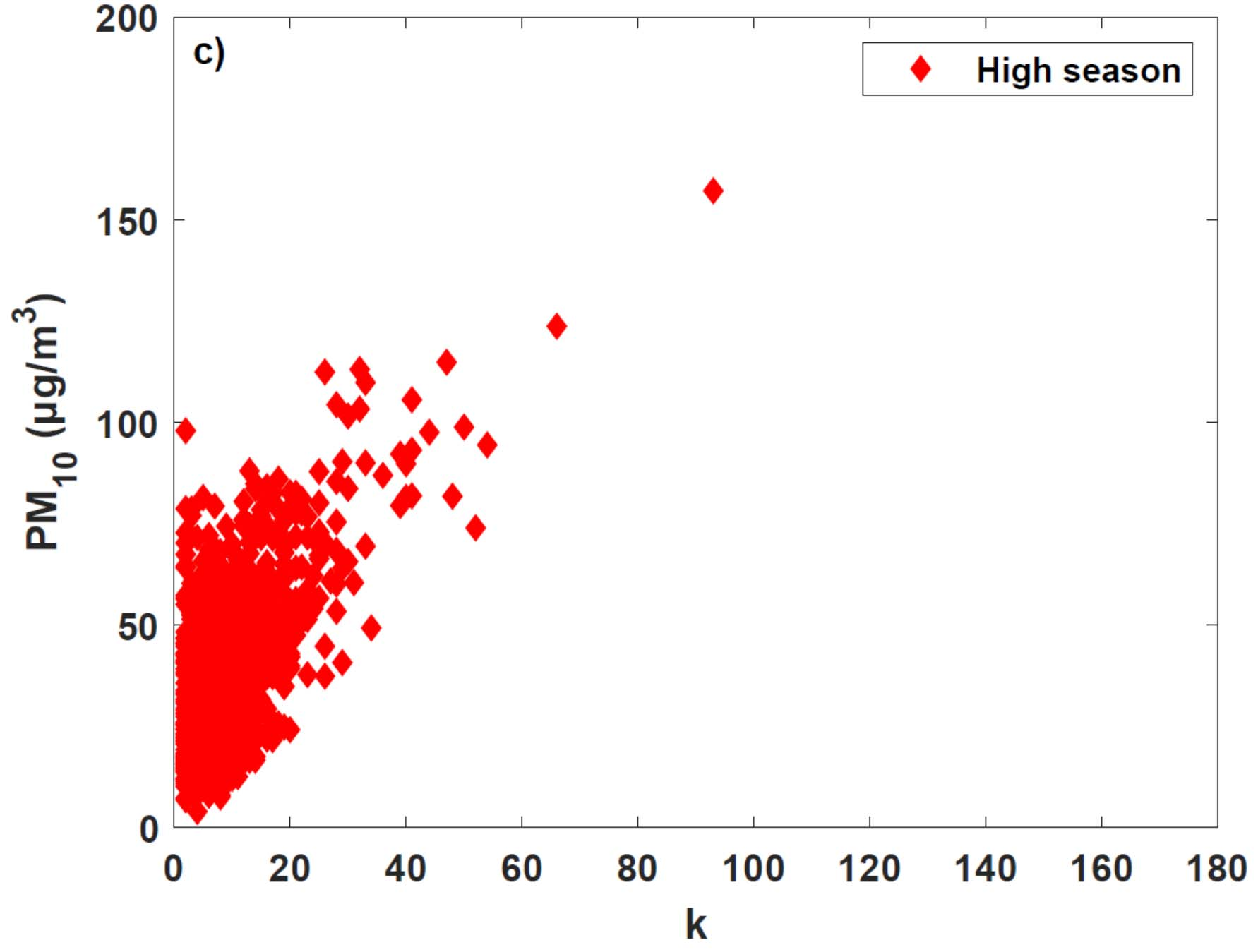}
\caption{\label{resTimesDeg} Relationship between $PM_{10}$  time series values and their degrees from VG frame for (a) all data, (b) low dust season (October to April) and (c) high dust season (May to September).}
\end{figure}

	From Rényi spectrum, several parameters can be obtained. First of all, the capacity or “box-counting” dimension $D_{0}$ related to the fractal object, i.e. how many boxes are needed in order to have it covered. In VG frame, a high extension would mean a higher degree, since the maximum covering of the system would be a complete graph ($K_n$) whose degree are maximum. As it could be expected, $D_0$ values for low and high dust seasons are consistent with $\bar{k}$. Intuitively, one could make the assumption that $D_0$ value for the “overall” would be between $D_0$ values of low and high dust seasons as $\gamma$ exponent previously found in section \ref{resultDeg}. Contrary to $\gamma$ which is based on data weight, $D_0$ is related to time series dynamics. “Overall” period highlights the seasonal variation of $PM_{10}$ data over 11 years while low and high dust seasons time series are reconstructed signals in order to take into account a specific period over 11 years. This may have an influence on fractal dimension behavior. In order to assess the relationship between the fractal object and the degree distribution, the values of $PM_{10}$ concentration are plotted against the degree in VG. This procedure (v-k plot) is frequently used to illustrate if highest degrees (hubs) are related to the largest concentrations \citep{pierini2012}. In Figure \ref{resTimesDeg}, one can observed that highest degrees correspond basically to the largest concentrations, specially in the high season ($k > 40$), since the accumulation of points in the low season is more equally distributed along the degree. Nonetheless, in both cases one could make the assumption that high concentrations will be reflected by the hubs in the complex network. This behavior was previously observe by \cite{carmona2019b} for tropospheric ozone time series in Cadiz, Spain. It is interesting to observe that v-k plot for “overall” is not just the sum of v-k plots for low and high dust seasons. As an example, 157.2 and 164.4 $\mu g/m^3$ respectively correspond to $k =$ 126 and 129 in “overall”. On the other hand, $k =$ 93 for 157.2 $\mu g/m^3$ in high season and $k =$ 152 for 164.4 $\mu g/m^3$ in low season. For the same concentration values, the degree can differ according to time series layout. All these results clearly highlight the fact that VG method takes into account $PM_{10}$ dynamics.
	
	Thereafter, the entropy dimension $D_{1}$ is introduced. The latter is related to the uniformity of the system. In order to quantify uniformity degree, the difference between $D_{0}$ and $D_{1}$ is computed. The greater $D_{0} - D_{1}$, the less uniform it is. In VG frame, greater uniformity means less difference between the degrees of the sample, i.e. a decreased of standard deviation for the degree distribution. Table \ref{VGresult} shows that the most uniform period is the high dust season because it exhibits minimum values for $D_{0} - D_{1}$ and $\sigma_k$. As illustrated in Figure \ref{signal}, one can observed in the middle of each year several peaks linked to African dust \citep{euphrasie2020}. Overall, these peaks seem more homogeneous between May and September due to the continuous alternation between African easterly waves and dust outbreaks \citep{prospero1981, karyampudi1999} which generate a repetitive pattern in $PM_{10}$ behavior. However, from October to April, some peaks with strong intensities can also be noted. These peaks can be caused by sporadic episodes of African dust transported by a particular circulation of air masses during Spring \citep{jury2017} or by extreme events as the eruption of Soufrière on Montserrat in February 2010 \citep{plocoste2019b}. This is the reason why they are called “high” and “low” seasons for African dust in literature \citep{prospero2014, velasco2018, plocoste2020} because depending on the synoptic conditions, dust outbreak can occur between October and April.
	
	 The last parameter estimated by Rényi spectrum is the correlation dimension $D_{2}$. This quantity is related to the recurrence in the series. In order to estimate recurrence degree, the difference between $D_{0}$ and $D_{2}$ is computed. The greater $D_{0} - D_{2}$, the higher the probability of finding the same value in the time series. Table \ref{VGresult} indicates that during the low dust season, recurrence degree is higher because it exhibits the maximum value of $D_{0} - D_{2}$. Based on $D_{0} - D_{1}$ discussion, this result make sense because African dust is less frequent during that period. Indeed, between October and April, the main $PM_{10}$ sources are marine aerosols and anthropogenic activity which composed the background atmosphere \citep{clergue2015, rastelli2017}. \cite{euphrasie2017} study shows that this background atmosphere is roughly 20 $\mu g/m^3$. Due to Guadeloupe flat topography in the center of the island and its weak anthropogenic activity \citep{plocoste2014, plocoste2018}, the diurnal variation of $PM_{10}$ concentrations is not significant as other megacities worldwide \citep{kunzli2000, mijic2009, sansuddin2011, xi2013, ozel2015, he2016, momtazan2019}. Thus, $PM_{10}$ values highlight a more persistence behavior between October and April, i.e. more predictable.

\subsubsection{Singularity spectrum}
\label{resultSing}

	To investigate the multifractal properties of $PM_{10}$ time series in VG frame through singularity spectrum, the \cite{chhabra1989} procedure is firstly performed. Here, the quantities $\sum\,\beta_i(q,r)\,ln[\beta_i(q,r)]$ and $\sum\,\beta_i(q,r)\,ln[P_i(r)]$ have been plotted against $ln\,r$ respectively for computed $f(\alpha)$ and $\alpha$. In order to maintain the scale used for Rényi spectrum, the same range of linear regression has been chosen to construct the singularity spectrum, i.e. $0 \leqslant ln\,r \leqslant 2$. The procedure to build singularity spectrum are widely described in the literature \citep{kelty2013, mali2018, carmona2019b}. In multifractal spectrum layout, the left side of $f(\alpha)$ spectrum is associated to $q > 0$ and it filters out the large fluctuations, whereas the right side of the spectrum is associated to $q < 0$ and it corresponds to small noise-like fluctuations. In VG theory, $q > 0$ is related to the highest values corresponding to the greatest degrees (so-called hubs) while $q < 0$ is related to the lowest values corresponding to the lowest degrees. 
		
\begin{figure}[h!]
\centering
\includegraphics[scale=0.45]{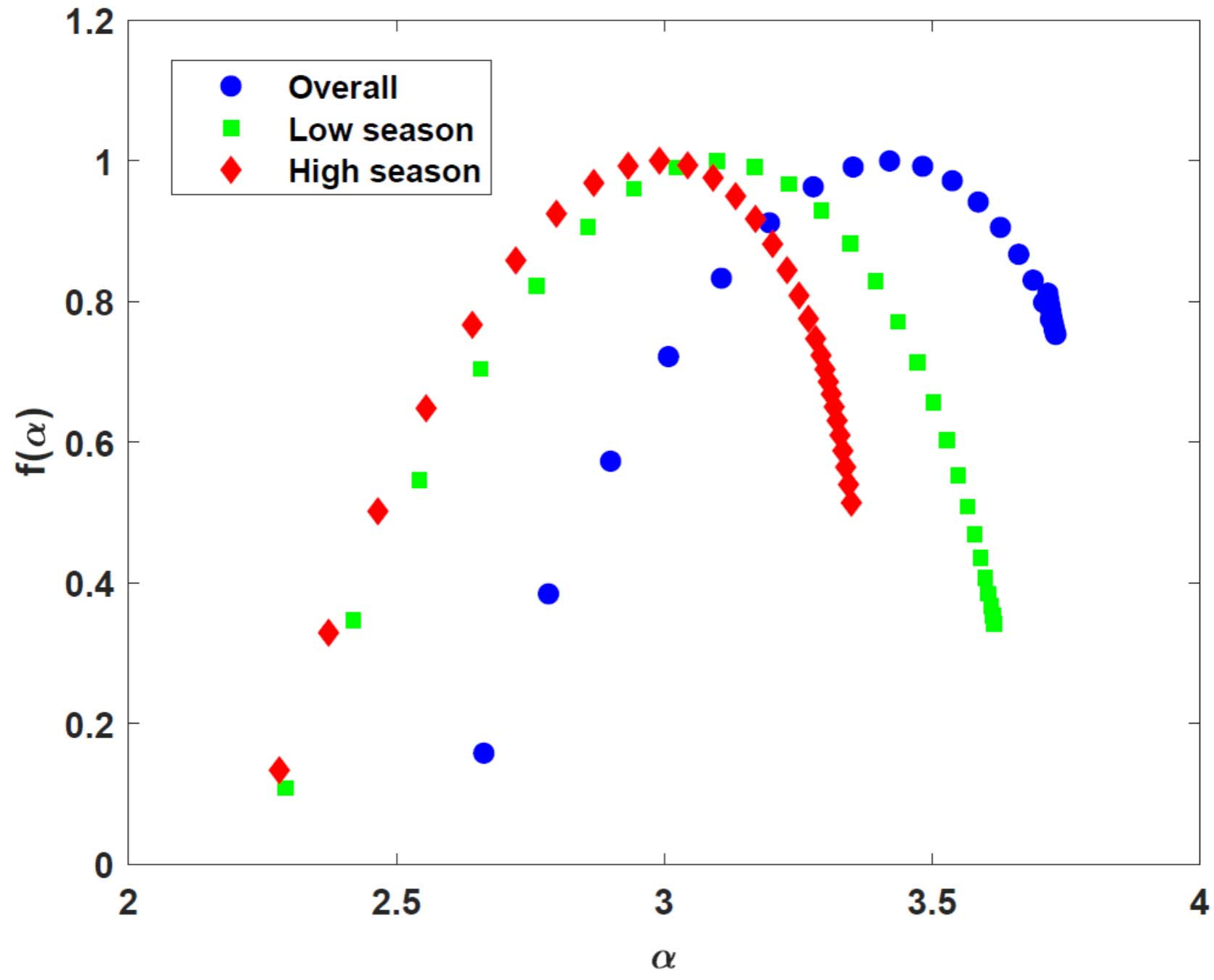}
\caption{\label{resSing} Singularity spectra for all data, low dust season (October to April) and high dust season (May to September).}
\end{figure}

	Figure \ref{resSing} shows the singularity spectra obtained where $f(\alpha)$ are plotted against $\alpha$. From Rényi spectrum, we found that the low dust season time series has a higher multifractality degree. It is in accordance with this singularity spectra, as the width of the curves is also related to this multifractal behavior. This is because the wider the spectrum, the larger will be the range of fractal exponents required to describe the signal. This is translated into a richer structure in the series. Again, it seems that the high dust season is marked by a less multifractal dynamics compared to the low dust one, i.e. $W_{High} < W_{Low}$ (see Table \ref{VGresult}).
	  
	Overall, one can notice that spectra are asymmetrical in shape with a left skewed distribution. This indicates dominance of large $\alpha$ values and therefore, a multifractal nature of large fluctuations. As it can be seen, the hubs are less likely and less homogeneous, as their $\alpha$ values differ more from each other. On the other hand, we observe that small noise-like fluctuations are more probable and homogeneous. By comparing both seasons, the left tail of the low season exhibits what could be expected, i.e. the large fluctuations are less probable ($f(\alpha)$ is lower in the left of low season). These results are consistent with what has been observed previously in the literature with high $PM_{10}$ values due to African dust from May to September, i.e. the high dust season \citep{prospero2014, velasco2018, euphrasie2020, plocoste2020a}.

\subsection{Centrality analysis}
\label{resultdMulti}

	In VG theory, another way to retrieve information from a given complex network is to assess nodes importance through centrality measures. Here, the monthly dynamics of $PM_{10}$ time series is investigated.
	
\begin{figure}[h!]
\centering
\includegraphics[scale=0.45]{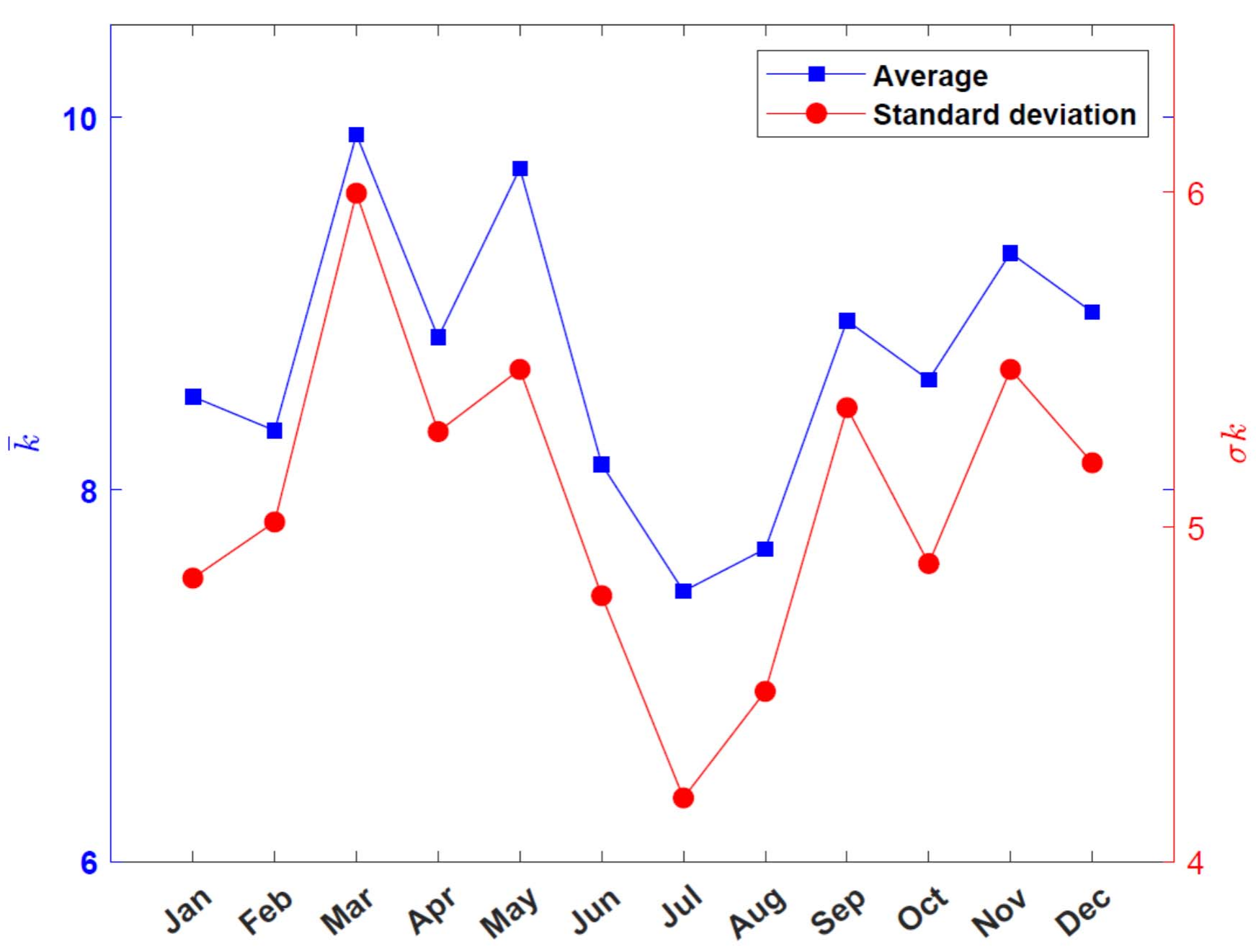}
\caption{\label{resCentDeg} Computed average degree and standard deviation from the degree distribution of each month over 11 years. Each monthly value is the average of the computed ones over 11 years.}
\end{figure}

	Figure \ref{resCentDeg} shows the monthly behavior of average degree ($\bar{k}$) and standard deviation ($\sigma_k$) from degree distribution over 11 years. Firstly, one can observe that the shape of both curves evolves simultaneously along the month. A trend emerges clearly with minimum values between May and September for both curves. In other words, during the high dust season a decay of hubs is observed. An opposite behavior was highlighted by \cite{carmona2019a} for ozone time series study in Cadiz, Spain. In their cases, the degree dropped during the months where high concentrations (hubs) were less significant. This difference in behavior may be due to many factors. Firstly, the nature of the pollutant. Contrary to ozone which is a secondary pollutant formed by photochemical reactions involving nitrogen oxides, carbon monoxyde and volatile organic compounds with solar energy \citep{blacet1952, seguel2012}; $PM_{10}$ is a primary pollutant principally linked to anthropogenic activity, marine aerosols and African dust in insular context \citep{plocoste2017, euphrasie2020}. Secondly, the scale of sources. Traditionally, ozone is mainly generated by the reaction between local precursors and solar radiation. Consequently a strong diurnal behavior is observed in ozone concentrations \citep{carmona2019a}. In Guadeloupe, the local emitters of $PM_{10}$ exhibit low concentrations, i.e. $\approx$ 20 $\mu g/m^3$ \citep{euphrasie2017}. This is the reason why high values of $PM_{10}$ are mainly attributed to African dust, i.e. large scale source \citep{plocoste2020b}. The diurnal magnitude of $PM_{10}$ values in Guadeloupe is weaker than megacities which have strong anthropogenic pollution (cars, factories etc). Thirdly, city configuration. In Guadeloupe, most of buildings are four stories high or less \citep{plocoste2014} contrary to megacities where high-rise buildings may influence wind circulation and promote pollutants accumulation in the surface layer. Furthermore, the permanent trade winds may increase pollutants dispersion and decrease risk of stagnation of pollutants in the atmospheric boundary layer.

\begin{figure}[h!]
\centering
\includegraphics[scale=0.45]{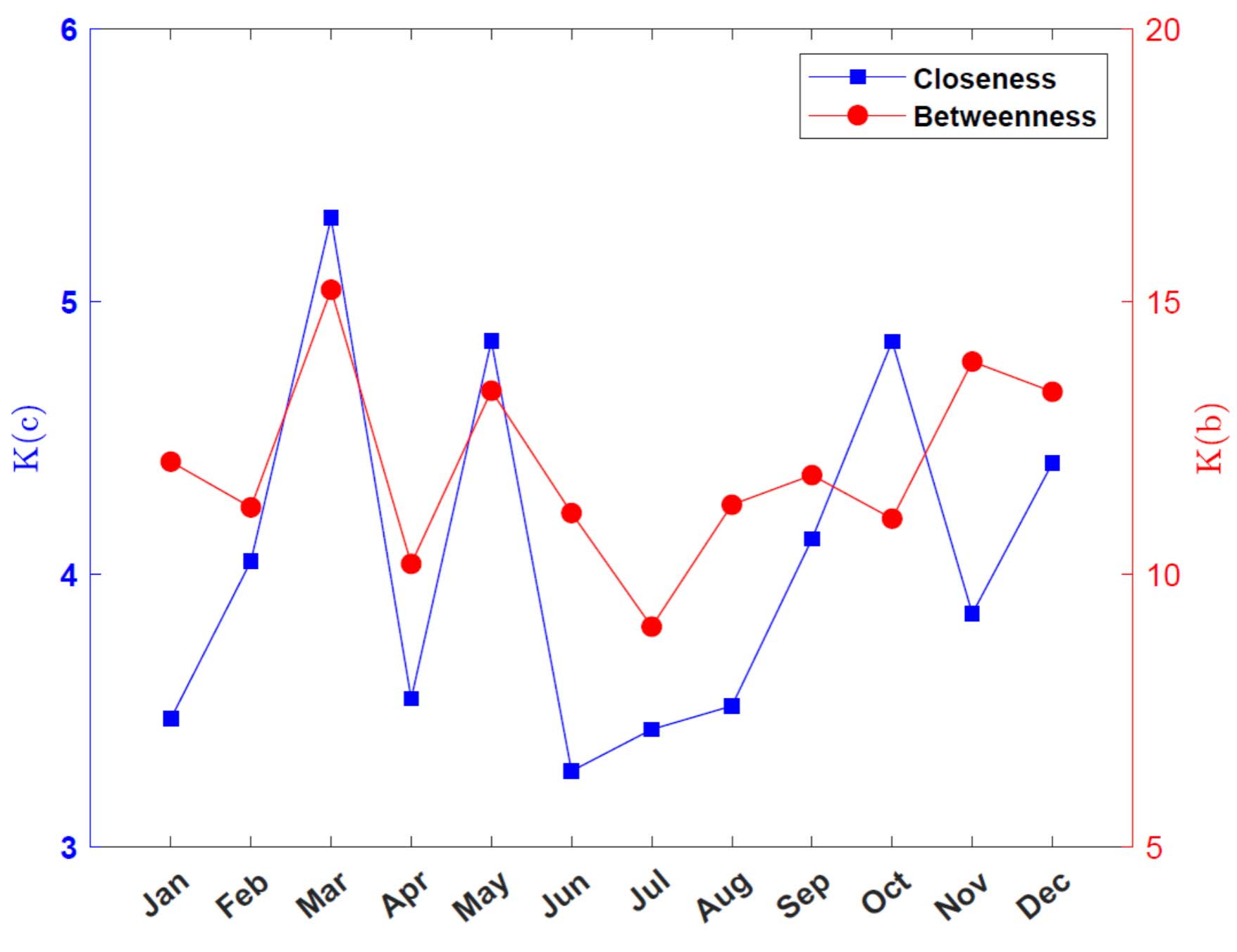}
\caption{\label{resCentSP} Computed average kurtosis from closeness and betweenness of each month over 11 years. Each monthly value is the kurtosis average of the computed ones over 11 years.}
\end{figure}

	Here, we also study $PM_{10}$ dynamics according to both centrality measures related to SP, i.e. closeness and betweenness. For these both quantities, the average kurtosis seems more sensitive to $PM_{10}$ dynamics than average, mediane and skewness. Figure \ref{resCentSP} depicts the monthly behavior of the average kurtosis values for both measures. Even if there are some fluctuations, a trend is clearly observed between May and September with a decrease of the average kurtosis values. The reasons previously mentioned to explain the decrease in degree centrality during the same period can also apply here.
	
	All these results clearly highlight that centrality measures are able to describe $PM_{10}$ time series dynamics. While a centred moment of order 2 is enough for degree centrality, a centred moment of order 4 is needed for centrality measures related to SP. This can be explained by the fact that degree centrality has an overview of the complex network while closeness and betweenness are more localized centrality measures, i.e. require a statistical parameter more sensitive to fluctuations.

\section{Conclusion}
\label{conclusion}

	In order to elaborate strategies and construct tools to predict dust events, it is fundamental to better understand particulate matter ($PM_{10}$) fluctuations. Here, the aim of this paper was to investigate $PM_{10}$ dynamics in the Caribbean basin according to African dust seasonality. To achieve this, 11 years of daily $PM_{10}$ database from Guadeloupe archipelago was analyzed using Visibility Graphs (VG).
		
	Firstly, the degree distribution $P(k)$ showed the self-similar properties of $PM_{10}$ data. Indeed, a scaling regime is found for all the studied periods highlighting the fractal nature of $PM_{10}$ time series. The exponents $\gamma$ computed from the slopes of $P(k)$ are estimated for $k \geq 10$ with $2.89 \leqslant \gamma \leqslant 3.17$.
		
	Thereafter, a multifractal analysis is performed through 2 approaches, i.e. the generalized fractal dimension and the singularity spectrum. Both methods pointed out a higher multifractality degree in the low dust season showing that the inhomogeneity of the distribution of probability measured on the overall fractal structure increases faster with the order of moments from October to April. In addition, multifractal parameters exhibited that the low dust season has the higher recurrence and the lower uniformity degrees. Authors believe that the continuous alternation between African easterly waves and dust outbreaks during the high dust season tends to homogenize the multifractal characteristics of $PM_{10}$ time series from May to September.
	
	Lastly, the monthly behavior of $PM_{10}$ time series is analyzed using the centrality measures, i.e. degree, closeness and betweenness. All the results showed that the centrality measures are sensitive to $PM_{10}$ dynamics through the year with a decay of the centrality values during the high dust season. While the average is enough for degree centrality analysis, the kurtosis is required for the others quantities related to the shortest path.   
	
	To conclude, all these results clearly showed that VG is a robust tool to describe times series properties. Given the many possibilities offered by complex networks, a wide range of subjects can be dealt with. In order to quantify a possible hub repulsion phenomenon in $PM_{10}$ time series, an analysis of edge repulsion force based on \cite{zhang2013} work should be carried out in a future study.

\section*{Acknowledgements}
The authors are very grateful to the anonymous reviewers for their valuable comments and constructive suggestions, which helped us to improve substantially the quality of the paper. The authors would like to thank Guadeloupe air quality network (Gwad'Air) for providing air quality data. A special thanks to Mr Sylvio Laventure for mapping assistance.  

\section*{Disclosure statement}

No potential conflict of interest was reported by the authors.

\section*{Funding}

The authors declare that they have not received any fund for the present paper. The paper is the sole work of the authors and is not a part/product of any project. 

\clearpage

\section*{References}
\bibliographystyle{model4-names}
\biboptions{authoryear}

\bibliography{VGPM10}

\end{document}